\documentclass[aip,graphicx]{revtex4-1}

\draft

\begin{document}

\title{Tetrads in low-energy weak interactions}

\author{ Alcides Garat$^{1}$ }

\address{1. Instituto de F\'{\i}sica, Facultad de Ciencias, Igu\'a 4225, esq. Mataojo, Montevideo, Uruguay.}
%\affiliation{1. Former Professor at Universidad de la Rep\'{u}blica, Instituto de F\'{\i}sica, Facultad de Ingenier\'{\i}a, J. Herrera y Reissig 565, 11300 Montevideo, Uruguay.}
\email[]{garat.alcides@gmail.com}
%Universidad de la Rep\'{u}blica,  Instituto de F\'{\i}sica, Facultad de Ciencias, Igu\'a 4225, esq. Mataojo, Montevideo, Uruguay.
\date{June 15th, 2006}

%\maketitle

\begin{abstract}

Tetrads are introduced in order to study the relationship between tetrad gauge states of spacetime and particle interactions, specially in weak processes at low energy. Through several examples like inverse Muon decay, elastic Neutrino-Electron scattering, it is explicitly shown how to assign to each vertex of the corresponding low-order Feynman diagram in a weak interaction, a particular set of tetrad vectors. The relationship between the tetrads associated to different vertices is exhibited explicitly to be generated by a $SU(2)$ local tetrad gauge transformation. We are establishing a direct link between standard gauge and tetrad gauge states of spacetime using the quantum field theories perturbative formulations.

\end{abstract}

\maketitle

\section{Introduction}
\label{introduction}

We are trying to understand the underlying symmetries of different field architectures, by showing explicitly the local geometrical structure of different kinds of groups of transformations of the Standard Model, specifically in their relationship to spacetime through specially defined tetrads. In references \cite{A,IWCP} we studied the local geometrical meaning of electromagnetic local gauge transformations. In references \cite{A2,A3,ASU3} we studied the local geometrical meaning of $SU(2) \times U(1)$ and $SU(3) \times SU(2) \times U(1)$ local groups of gauge transformations. Isomorphisms and homomorphisms were found that relate the standard model groups of local gauge transformations with new groups of local geometrical transformations in four-dimensional curved Lorentz spacetimes. These relationships can be explicitly displayed through the use of appropriately defined tetrads. It is the purpose of this work, to make use of already defined tetrads of different kinds \cite{A,IWCP,A2,A3,ASU3}, in order to briefly show in an explicit way, the invariance of the metric tensor associated to a low-energy weak interaction, under different kinds of transformations. For instance, the invariance under electromagnetic local gauge transformations, the invariance under $SU(2)$ local gauge transformations, the invariance under local gauge transformations of the spinor fields \cite{MK,GM}, etc. Since we are trying to ``geometrize'' the local gauge theories, it is interesting in its own right, to understand as well, the geometries that involve the standard fields associated with microparticle interactions. To that end, we introduce what we call ``tetrad Feynman calculus''. We are able to explicitly show how to build a tetrad associated to a Feynman low-order diagram in low-energy weak interactions. The massive weak interactions bosons must have an associated gravitational field as electrons, muons and neutrinos also do have and even though these gravitational fields might be weak, they possess the necessary geometrical structure that enables the local symmetries of the standard model to be realized in an explicit fashion as it was analyzed in previous manuscripts \cite{A,IWCP,A2,A3,ASU3}. In high energy interactions where virtual phenomena becomes relevant, a different approach is needed as we will discuss later on. We proceed to show how to assign a tetrad to each vertex, for instance in inverse Muon decay, and elastic Neutrino-Electron scattering. In the first two sections \ref{overview} and \ref{nonabeltetrads} we deal as an introduction with these tetrads defined in a general curved four-dimensional Lorentzian spacetime. These two sections and because of the construction nature of the new tetrads can also be identically developed in flat Minkowski spacetimes. In section \ref{spacetimefeynman} we will limit our analysis to flat spacetimes as an example compatible with the foundations of quantum field theories. As a general thought we strongly believe that the construction of tetrad fields, and metric tensors that explicitly display the local symmetries of microparticle interactions, are hinting us over a possible relationship or link, between General Relativity and Quantum Theories. In both subsections \ref{invmuon}-\ref{elasticneutrino} of section \ref{spacetimefeynman} we also demonstrate that it is possible to transform the tetrad associated to a vertex in a particular diagram to the tetrad assigned to another vertex in the same Feynman diagram through a local $SU(2)$ tetrad gauge transformation at the same spacetime point. It is clear that we envisage the spacetime of the interaction process as a common spacetime for all participating objects in the region of interaction even though in this first manuscript on the subject we are addressing the processes on a flat spacetime background. Throughout the paper we use the conventions of references \cite{A,A2,MW}. In particular we use a metric with sign conventions -+++. The only difference in notation with \cite{MW} will be that we will call our geometrized electromagnetic potential $A^{\alpha}$, where $f_{\mu\nu}=A_{\nu ;\mu} - A_{\mu ;\nu}$ is the geometrized electromagnetic field $f_{\mu\nu}= (G^{1/2} / c^2) \: F_{\mu\nu}$. Analogously, $f^{k}_{\mu\nu}$ are the geometrized Yang-Mills field components, $f^{k}_{\mu\nu}= (G^{1/2} / c^2) \: F^{k}_{\mu\nu}$.

\section{Overview of new tetrads and symmetries for the Abelian case}
\label{overview}

%We will take the risk of becoming redundant, by reiterating notions presented in this overview section, that have already been introduced in previous papers \cite{dsmg}$^{,}$\cite{gaugeinvmeth}, with the focussed objective of making this work self-contained.
The new tetrads have been designed on the basis of existence of antisymmetric second rank tensors. In the particular case were Abelian non-null electromagnetic fields are present in spacetime in addition to curvature or a gravitational field, or even when spacetime is flat the new method involves a local duality rotation of these gauge fields. We then proceed to introduce at every point in spacetime a duality rotation by an angle $-\alpha$ that transforms a non-null electromagnetic field $f_{\mu\nu}$ into an extremal field $\xi_{\mu\nu}$,

\begin{equation}
\xi_{\mu\nu} = e^{-\ast \alpha} f_{\mu\nu}\ = \cos(\alpha)\:f_{\mu\nu} - \sin(\alpha)\:\ast f_{\mu\nu},\label{drefn}
\end{equation}

where  $\ast f_{\mu\nu}={1 \over 2}\:\epsilon_{\mu\nu\sigma\tau}\:f^{\sigma\tau}$ is the dual tensor of $f_{\mu\nu}$. The local scalar $\alpha$ is the complexion of the electromagnetic field and it is a local gauge invariant quantity. Extremal fields are essentially electric fields and they satisfy,

\begin{equation}
\xi_{\mu\nu} \ast \xi^{\mu\nu}= 0\ . \label{ia}
\end{equation}

Equation (\ref{ia}) is imposed as a condition on (\ref{drefn}) and then we find the expression for the complexion that results in $\tan(2\alpha) = - f_{\mu\nu}\:\ast f^{\mu\nu} / f_{\lambda\rho}\:f^{\lambda\rho}$. We can also prove just using identities valid in four-dimensional Lorentzian spacetimes that the condition (\ref{ia}) can be rewritten as $\xi_{\alpha\mu}\:\ast \xi^{\mu\nu} = 0$. In order to prove this equivalence between conditions we will need an identity \cite{MW} valid for two second rank antisymmetric fields in a four-dimensional Lorentzian spacetime. This identity is given by,

\begin{eqnarray}
A_{\mu\alpha}\:B^{\nu\alpha} -
\ast B_{\mu\alpha}\: \ast A^{\nu\alpha} &=& \frac{1}{2}
\: \delta_{\mu}^{\:\:\:\nu}\: A_{\alpha\beta}\:B^{\alpha\beta}  \ ,\label{ib}
\end{eqnarray}

When this identity (\ref{ib}) is considered for the case $A_{\mu\alpha} = \xi_{\mu\alpha}$ and $B^{\nu\alpha} = \xi^{\nu\alpha}$ we obtain,

\begin{eqnarray}
\xi_{\mu\alpha}\:\xi^{\nu\alpha} -
\ast \xi_{\mu\alpha}\: \ast \xi^{\nu\alpha} &=& \frac{1}{2}
\: \delta_{\mu}^{\:\:\:\nu}\ Q \ ,\label{ic}
\end{eqnarray}

where $Q=\xi_{\mu\nu}\:\xi^{\mu\nu}=-\sqrt{T_{\mu\nu}T^{\mu\nu}}$ according to equations (39) in reference \cite{MW}. $Q$ is assumed not to be zero, because we are dealing with non-null electromagnetic fields. It can be proved that condition (\ref{ia}) plus the general identity (\ref{ib}), when applied to the case $A_{\mu\alpha} = \xi_{\mu\alpha}$ and $B^{\nu\alpha} = \ast \xi^{\nu\alpha}$ provides the equivalent condition to (\ref{ia}),

\begin{eqnarray}
\xi_{\alpha\mu}\:\ast \xi^{\mu\nu} &=& 0\ ,\label{id}
\end{eqnarray}

which is equation (64) in reference \cite{MW}. In geometrodynamics, the Einstein-Maxwell equations,

\begin{eqnarray}
R_{\mu\nu} &=& f_{\mu\lambda}\:\:f_{\nu}^{\:\:\:\lambda}
+ \ast f_{\mu\lambda}\:\ast f_{\nu}^{\:\:\:\lambda} \label{L0}  \\
f^{\mu\nu}_{\:\:\:\:\:;\nu} &=& 0 \label{L1} \\
\ast f^{\mu\nu}_{\:\:\:\:\:;\nu} &=& 0 \ , \label{L2}
\end{eqnarray}

reveal the existence of two potential vector fields \cite{CF} $A_{\nu}$ and $\ast A_{\nu}$,

\begin{eqnarray}
f_{\mu\nu} &=& A_{\nu ;\mu} - A_{\mu ;\nu}\label{ERN} \\
\ast f_{\mu\nu} &=& \ast A_{\nu ;\mu} - \ast A_{\mu ;\nu} \ .\label{DERN}
\end{eqnarray}

The symbol $``;''$ stands for covariant derivative with respect to the metric tensor $g_{\mu\nu}$ and the star in $\ast A_{\nu}$ is just nomenclature, not the dual operator, with the meaning that $\ast A_{\nu ;\mu} = (\ast A_{\nu})_{;\mu}$. The duality rotation given by equation (59) in\cite{MW} $f_{\mu\nu} = \xi_{\mu\nu} \: \cos\alpha + \ast\xi_{\mu\nu} \: \sin\alpha$, enables us to reexpress the stress-energy tensor in terms of the extremal field,

\begin{equation}
T_{\mu\nu}=\xi_{\mu\lambda}\:\:\xi_{\nu}^{\:\:\:\lambda}
+ \ast \xi_{\mu\lambda}\:\ast \xi_{\nu}^{\:\:\:\lambda}\ .\label{TEMDRN}
\end{equation}

It is now the right time to introduce the new tetrads that will diagonalize locally and covariantly the stress-energy tensor (\ref{TEMDRN}).

\begin{eqnarray}
U^{\alpha} &=& \xi^{\alpha\lambda}\:\xi_{\rho\lambda}\:X^{\rho} \:
/ \: (\: \sqrt{-Q/2} \: \sqrt{X_{\mu} \ \xi^{\mu\sigma} \
\xi_{\nu\sigma} \ X^{\nu}}\:) \label{UN}\\
V^{\alpha} &=& \xi^{\alpha\lambda}\:X_{\lambda} \:
/ \: (\:\sqrt{X_{\mu} \ \xi^{\mu\sigma} \
\xi_{\nu\sigma} \ X^{\nu}}\:) \label{VN}\\
Z^{\alpha} &=& \ast \xi^{\alpha\lambda} \: Y_{\lambda} \:
/ \: (\:\sqrt{Y_{\mu}  \ast \xi^{\mu\sigma}
\ast \xi_{\nu\sigma}  Y^{\nu}}\:)
\label{ZN}\\
W^{\alpha} &=& \ast \xi^{\alpha\lambda}\: \ast \xi_{\rho\lambda}
\:Y^{\rho} \: / \: (\:\sqrt{-Q/2} \: \sqrt{Y_{\mu}
\ast \xi^{\mu\sigma} \ast \xi_{\nu\sigma} Y^{\nu}}\:) \ .
\label{WN}
\end{eqnarray}

With all these elements put together, particularly equations (\ref{ic}-\ref{id}) it becomes trivial to prove that the tetrad (\ref{UN}-\ref{WN}) is orthonormal and diagonalizes \cite{A} the stress-energy tensor (\ref{TEMDRN}). The vectors (\ref{UN}-\ref{VN}) with eigenvalue $Q/2$ and the vectors (\ref{ZN}-\ref{WN}) with eigenvalue $-Q/2$. At every point in spacetime the timelike and one spacelike vectors that for some geometries like Reissner-Nordstr\"{o}m are (\ref{UN}-\ref{VN}) generate a plane that we called blade one \cite{A}$^{,}$\cite{SCH}. The other two spacelike vectors (\ref{ZN}-\ref{WN}) generate a local orthogonal plane that we called blade two. These vectors are constructed with the local extremal field \cite{MW} (\ref{drefn}), its dual, the very metric tensor and a pair of vector fields $X^{\alpha}$ and $Y^{\alpha}$ that represent a generic gauge choice as long as the tetrad vectors do not become trivial. We are aware then, that we still have to introduce the vectors $X^{\mu}$ and $Y^{\mu}$. Let us introduce some names. The tetrad vectors have two essential structure components. For instance in vector $U^{\alpha}$ there are two main structures. First, the skeleton, in this case $\xi^{\alpha\lambda}\:\xi_{\rho\lambda}$, and second, the gauge vector $X^{\rho}$. These do not include the normalization factor $1 / \: (\: \sqrt{-Q/2} \: \sqrt{X_{\mu} \ \xi^{\mu\sigma} \ \xi_{\nu\sigma} \ X^{\nu}}\:)$. The gauge vectors must be anything that do not make the tetrad vectors trivial and we mean by this that the tetrad (\ref{UN}-\ref{WN}) diagonalizes the stress-energy tensor for any non-trivial gauge vectors $X^{\mu}$ and $Y^{\mu}$. It is then possible to make different choices for $X^{\mu}$ and $Y^{\mu}$. The potential vector fields introduced in equations (\ref{ERN}-\ref{DERN}) represent a possible choice in geometrodynamics for the vectors $X^{\alpha}=A^{\alpha}$ and $Y^{\alpha}=\ast A^{\alpha}$. We do not mean that the two vector fields have independence from each other, it is just a convenient choice. With this particular choice for the two gauge vector fields we can define then,

\begin{eqnarray}
U^{\alpha} &=& \xi^{\alpha\lambda}\:\xi_{\rho\lambda}\:A^{\rho} \:
/ \: (\: \sqrt{-Q/2} \: \sqrt{A_{\mu} \ \xi^{\mu\sigma} \
\xi_{\nu\sigma} \ A^{\nu}}\:) \label{UON}\\
V^{\alpha} &=& \xi^{\alpha\lambda}\:A_{\lambda} \:
/ \: (\:\sqrt{A_{\mu} \ \xi^{\mu\sigma} \
\xi_{\nu\sigma} \ A^{\nu}}\:) \label{VON}\\
Z^{\alpha} &=& \ast \xi^{\alpha\lambda} \: \ast A_{\lambda} \:
/ \: (\:\sqrt{\ast A_{\mu}  \ast \xi^{\mu\sigma}
\ast \xi_{\nu\sigma}  \ast A^{\nu}}\:)
\label{ZON}\\
W^{\alpha} &=& \ast \xi^{\alpha\lambda}\: \ast \xi_{\rho\lambda}
\:\ast A^{\rho} \: / \: (\:\sqrt{-Q/2} \: \sqrt{\ast A_{\mu}
\ast \xi^{\mu\sigma} \ast \xi_{\nu\sigma} \ast A^{\nu}}\:) \ ,
\label{WON}
\end{eqnarray}

where the four vectors (\ref{UON}-\ref{WON}) satisfy the following algebraic properties,

\begin{equation}
-U^{\alpha}\:U_{\alpha}=V^{\alpha}\:V_{\alpha}=Z^{\alpha}\:Z_{\alpha}=W^{\alpha}\:W_{\alpha}=1 \ .\label{TSPAUXN}
\end{equation}

Using the equations (\ref{ic}-\ref{id}) it is simple to prove that (\ref{UON}-\ref{WON}) are orthogonal vectors. We think then about local electromagnetic gauge transformations. We notice that we can interpret the independent local gauge transformations of the vector potentials introduced in equations (\ref{ERN}-\ref{DERN}), that is,  $A_{\alpha} \rightarrow A_{\alpha} + \Lambda_{,\alpha}$ and $\ast A_{\alpha} \rightarrow \ast A_{\alpha} + \ast \Lambda_{,\alpha}$ as new choices for the two gauge vector fields $X^{\mu}$ and $Y^{\mu}$. The first local gauge transformation leaves $f_{\mu\nu}$ invariant and the second one leaves $\ast f_{\mu\nu}$ invariant, as long as the functions $\Lambda$ and $\ast \Lambda$ are scalars. According to Schouten, for non-null electromagnetic fields in Einstein-Maxwell spacetimes there is a two-bladed or two-plane structure \cite{SCH} at every point in spacetime. These blades are the planes determined by the pairs ($U^{\alpha}, V^{\alpha}$) and ($Z^{\alpha}, W^{\alpha}$). In manuscript \cite{A} it was demonstrated that the transformation $A_{\alpha} \rightarrow A_{\alpha} + \Lambda_{,\alpha}$ generates a Lorentz transformation (except for one discrete reflection) of the tetrad vectors ($U^{\alpha}, V^{\alpha}$) into ($\tilde{U}^{\alpha}, \tilde{V}^{\alpha}$) in such a way that these ``rotated'' vectors ($\tilde{U}^{\alpha}, \tilde{V}^{\alpha}$) remain in the plane or blade one generated by ($U^{\alpha}, V^{\alpha}$). In the same reference \cite{A} it was also proven that the transformation $\ast A_{\alpha} \rightarrow \ast A_{\alpha} + \ast \Lambda_{,\alpha}$ generates a ``rotation'' of the tetrad vectors ($Z^{\alpha}, W^{\alpha}$) into ($\tilde{Z}^{\alpha}, \tilde{W}^{\alpha}$) such that these ``rotated'' vectors ($\tilde{Z}^{\alpha}, \tilde{W}^{\alpha}$) remain in the plane or blade two generated by ($Z^{\alpha}, W^{\alpha}$). In manuscript \cite{A} it was demonstrated that the group of local electromagnetic gauge transformations is isomorphic to the group LB1 of boosts plus two discrete transformations on blade one, and independently to LB2, the group of spatial rotations on blade two. Equations like

\begin{eqnarray}
U^{\alpha}_{(\phi)}  &=& \cosh(\phi)\: U^{\alpha} +  \sinh(\phi)\: V^{\alpha} \label{UTN} \\
V^{\alpha}_{(\phi)} &=& \sinh(\phi)\: U^{\alpha} +  \cosh(\phi)\: V^{\alpha} \label{VTN} \ ,
\end{eqnarray}

on the local plane one represent a local electromagnetic gauge transformation of the vectors $(U^{\alpha}, V^{\alpha})$. The transformation of the two vectors $(U^{\alpha},\:V^{\alpha})$ on blade one, given in (\ref{UON}-\ref{VON}) by the ``angle'' $\phi$ in (\ref{UTN}-\ref{VTN}) is a proper transformation, that is, a boost. For discrete improper transformations the result follows the same lines, see reference \cite{A}. Analogously on the local plane two,

\begin{eqnarray}
Z^{\alpha}_{(\varphi)}  &=& \cos(\varphi)\: Z^{\alpha} -  \sin(\varphi)\: W^{\alpha} \label{ZTN} \\
W^{\alpha}_{(\varphi)}  &=& \sin(\varphi)\: Z^{\alpha} +  \cos(\varphi)\: W^{\alpha} \label{WTN} \ .
\end{eqnarray}

Equations (\ref{ZTN}-\ref{WTN}) represent a local electromagnetic gauge transformation of the vectors $(Z^{\alpha}, W^{\alpha})$, the transformation of the two tetrad vectors $(Z^{\alpha},\:W^{\alpha})$ on blade two, given in (\ref{ZON}-\ref{WON}), by the ``angle'' $\varphi$. It is straightforward to check that the equalities $U^{[\alpha}_{(\phi)}\:V^{\beta]}_{(\phi)} = U^{[\alpha}\:V^{\beta]}$ and $Z^{[\alpha}_{(\varphi)}\:W^{\beta]}_{(\varphi)} = Z^{[\alpha}\:W^{\beta]}$ are true. These equalities mean that these antisymmetric tetrad objects are gauge invariant. In the Abelian case it was proved that the local group of electromagnetic gauge transformations is isomorphic to both the local groups LB1 and LB2, separately, independently. LB1 on the local plane one is a group composed by the tetrad boosts $SO(1,1)$ and two different kinds of discrete transformations. One of the discrete transformations is the full inversion or minus the identity two by two. The other discrete transformation is not Lorentzian \cite{A} because it is a reflection or flip, a two by two matrix with zeroes in the diagonal and ones off-diagonal. LB2 on plane two is the group of spatial tetrad rotations on this plane, that is $SO(2)$. It is worth reminding ourselves about a point in mathematical language that could be loose or inaccurate but nonetheless immediate to understand. With the purpose of exemplifying we can mention the isomorphisms in the Abelian case \cite{A} between the local group of electromagnetic gauge transformations and the local groups of tetrad transformations LB1 and LB2, separately, independently. The isomorphisms strictly speaking are homomorphisms between the local algebra of scalars associated to the local group of electromagnetic gauge transformations and the local groups LB1 and LB2, independently. We know that between the local algebra of scalars and the local group of electromagnetic gauge transformations there is a homomorphism, a homomorphism between the real numbers $\mathcal{R}$ that is, the algebra of local scalars associated to the local group of electromagnetic gauge transformations and $U(1)$, that is, the local group of electromagnetic gauge transformations. We give this relationship as implicitly understood even though we talk about isomorphisms between the local group of electromagnetic gauge transformations and the local groups of tetrad transformations LB1 and LB2, separately, independently.

We must also stress that the local transformations (\ref{UTN}-\ref{VTN}) are not imposed local boosts on the vectors that define the local plane one. They are the result of local gauge transformations of the vectors ($U^{\alpha}, V^{\alpha}$). For example, from reference \cite{A} a particular boost after the gauge transformation would look like,

\begin{eqnarray}
{\tilde{V}_{(1)}^{\alpha}
\over \sqrt{-\tilde{V}_{(1)}^{\beta}\:\tilde{V}_{(1)\beta}}}&=&
{(1+C) \over \sqrt{(1+C)^2-D^2}}
\:{V_{(1)}^{\alpha} \over \sqrt{-V_{(1)}^{\beta}\:V_{(1)\beta}}}+
{D \over \sqrt{(1+C)^2-D^2}}
\:{V_{(2)}^{\alpha} \over \sqrt{V_{(2)}^{\beta}\:V_{(2)\beta}}}\label{TN1N}\\
{\tilde{V}_{(2)}^{\alpha}
\over \sqrt{\tilde{V}_{(2)}^{\beta}\:\tilde{V}_{(2)\beta}}}&=&
{D \over \sqrt{(1+C)^2-D^2}}
\:{V_{(1)}^{\alpha} \over \sqrt{-V_{(1)}^{\beta}\:V_{(1)\beta}}} +
{(1+C) \over \sqrt{(1+C)^2-D^2}}
\:{V_{(2)}^{\alpha} \over \sqrt{V_{(2)}^{\beta}\:V_{(2)\beta}}}\ .
\label{TN2N}
\end{eqnarray}

In equations (\ref{TN1N}-\ref{TN2N}) the following notation has been used, $C = (-Q/2)\:V_{(1)\sigma}\:\Lambda^{\sigma} / (\:V_{(2)\beta}\:V_{(2)}^{\beta}\:)$, $D = (-Q/2)\:V_{(2)\sigma}\:\Lambda^{\sigma} / (\:V_{(1)\beta}\:V_{(1)}^{\beta}\:)$ and $[(1+C)^2-D^2]>0$ must be satisfied. The notation $\Lambda^{\alpha}$ has been used for $\Lambda^{,\alpha}$ where $\Lambda$ is the local scalar generating the local gauge transformation. $U^{\alpha} = {V_{(1)}^{\alpha} \over \sqrt{-V_{(1)}^{\beta}\:V_{(1)\beta}}}$ and $V^{\alpha} = {V_{(2)}^{\alpha} \over \sqrt{V_{(2)}^{\beta}\:V_{(2)\beta}}}$ according to the notation used in paper \cite{A},

\begin{eqnarray}
V_{(1)}^{\alpha} &=& \xi^{\alpha\lambda}\:\xi_{\rho\lambda}\:A^{\rho}
\label{V1A}\\
V_{(2)}^{\alpha} &=& \sqrt{-Q/2} \: \xi^{\alpha\lambda} \: A_{\lambda}
\label{V2A}\\
V_{(3)}^{\alpha} &=& \sqrt{-Q/2} \: \ast \xi^{\alpha\lambda}
\: \ast A_{\lambda}\label{V3A}\\
V_{(4)}^{\alpha} &=& \ast \xi^{\alpha\lambda}\: \ast \xi_{\rho\lambda}
\:\ast A^{\rho}\ .\label{V4A}
\end{eqnarray}

For the particular case when $1+C > 0$, the transformations (\ref{TN1N}-\ref{TN2N}) manifest that an electromagnetic gauge transformation on the vector field $A^{\alpha} \rightarrow A^{\alpha} + \Lambda^{\alpha}$, that leaves invariant the electromagnetic field $f_{\mu\nu}$, generates a boost transformation on the normalized tetrad vector fields $\left({V_{(1)}^{\alpha} \over \sqrt{-V_{(1)}^{\beta}\:V_{(1)\beta}}}, {V_{(2)}^{\alpha} \over \sqrt{V_{(2)}^{\beta}\:V_{(2)\beta}}}\right)$. In this case $\cosh(\phi) = {(1+C) \over \sqrt{(1+C)^2-D^2}}$, see also equation (\ref{UTN}). This was just one of the possible cases in LB1. Similar analysis for the vector transformations (\ref{ZTN}-\ref{WTN}) in the local plane two generated by ($Z^{\alpha}, W^{\alpha}$). See reference \cite{A} for the detailed analysis of all possible cases.

Back to our main line of work we can write the electromagnetic field in terms of these tetrad vectors,

\begin{equation}
f_{\alpha\beta} = -2\:\sqrt{-Q/2}\:\:\cos\alpha\:\:U_{[\alpha}\:V_{\beta]} +
2\:\sqrt{-Q/2}\:\:\sin\alpha\:\:Z_{[\alpha}\:W_{\beta]}\ .\label{EMF}
\end{equation}

Equation (\ref{EMF}) entails the maximum simplification in the expression of the electromagnetic field. The true degrees of freedom are the local scalars $\sqrt{-Q/2}$ and $\alpha$. We can also present both local degrees of freedom as $\sqrt{-Q/2}\:\cos\alpha$ and $\sqrt{-Q/2}\:\sin\alpha$. The object $U_{[\alpha}\:V_{\beta]}$ remains invariant \cite{A} under a ``rotation'' of the tetrad vectors $U^{\alpha}$ and $V^{\alpha}$ by a scalar angle $\phi$ like in (\ref{UTN}-\ref{VTN}) on blade one. This is the way in which local gauge invariance is manifested explicitly on this local plane. Analogous for discrete transformations on blade one. Similar analysis on blade two. A spatial ``rotation'' generated by a gauge transformation of the tetrad vectors $Z^{\alpha}$ and $W^{\alpha}$ through an ``angle'' $\varphi$ as in (\ref{ZTN}-\ref{WTN}), such that the object $Z_{[\alpha}\:W_{\beta]}$ remains invariant \cite{A}. All this formalism clearly provides a technique to maximally simplify the expression for the electromagnetic field strength as in equation (\ref{EMF}). It is block diagonalized automatically by the tetrad (\ref{UON}-\ref{WON}). This is not the case for the non-Abelian $SU(2)$ field strength. We do not have an automatic block diagonalization. To this purpose a new algorithm was developed in reference \cite{gaugeinvmeth}. In the next section we study the construction of tetrads similar to the Abelian case but in the non-Abelian environment.

\section{Overview of new tetrads and symmetries for the non-Abelian case}
\label{nonabeltetrads}

%We could also skip this section and just cite manuscripts \cite{A2,A3,ASU3}$^{,}$\cite{dsmg,gaugeinvmeth}. We choose not to, in order to be able to handle in the next section substantial and new concepts that require the constant reference of details that belong to this, once again, redundant account.

For the non-Abelian case we first would like to present a set of examples on how to build extremal fields that are locally invariant under $SU(2)$ gauge transformations. Let us remember that in the Abelian case in the tetrads (\ref{UN}-\ref{WN}) the only dependence on gauge came through the gauge vectors $X^{\mu}$ and $Y^{\mu}$. The tetrad skeletons as it was mentioned previously are local gauge invariants. The advantage of this method is that when we introduce local gauge transformations, the vectors that span the local plane one, do not leave this plane after the transformation and analogous in the local plane two. This fact implies in turn that the metric tensor, be non-flat or flat will not change and will remain invariant. It is then important to show explicitly that we can construct extremal fields invariant under both the Abelian and the non-Abelian gauge transformations. One example could be for instance given by,

\begin{equation}
\zeta_{\mu\nu} = \cos\beta \:\: f_{\mu\nu}-\sin\beta \:\: \ast f_{\mu\nu} \ ,\label{exsu2n}
\end{equation}

Following the Abelian pattern we can define the new complexion $\beta$, and to this end we will impose the new $SU(2)$ local invariant condition,

\begin{eqnarray}
Tr[\zeta_{\mu\nu}\:\ast \zeta^{\mu\nu}]=\zeta^{k}_{\mu\nu}\:\ast \zeta^{k\mu\nu} &=& 0\ ,\label{ccsu2n}
\end{eqnarray}

where the summation convention is applied on the internal index $k$. We are just using a generalized duality transformation, and defining through it, this new local scalar complexion $\beta$. Therefore, the complexion condition (\ref{ccsu2n}) is not an additional condition for the field strength. We simply introduced a possible generalization of the definition for the Abelian complexion, found through a new duality transformation as well. Then, we find the local $SU(2)$ invariant complexion $\beta$ to be,

\begin{eqnarray}
\tan(2\beta) = - f^{k}_{\mu\nu}\:\ast f^{k\mu\nu} / f^{p}_{\lambda\rho}\:f^{p\lambda\rho}\ ,\label{compksu2}
\end{eqnarray}

where once again the summation convention was applied on both $k$ and $p$. We can also consider gauge covariant derivatives since they will become useful in the ensuing analysis. For example, the gauge covariant derivatives of the three extremal field internal components,

\begin{eqnarray}
\zeta_{k\mu\nu\mid\rho} = \zeta_{k\mu\nu\, ; \, \rho} + g \: \epsilon_{klp}\: A_{l\rho}\:\zeta_{p\mu\nu}\ .\label{gcd}
\end{eqnarray}

where $\epsilon_{klp}$ is the completely skew-symmetric tensor in three dimensions with $\epsilon_{123} = 1$, with $g$ the coupling constant. As in the previous section \ref{overview} the symbol ``;'' stands for the usual covariant derivative associated with the metric tensor $g_{\mu\nu}$. Next we consider the Einstein-Maxwell-Yang-Mills vacuum field equations,

\begin{eqnarray}
R_{\mu\nu} &=& T^{(ym)}_{\mu\nu} + T^{(em)}_{\mu\nu}\label{eymen}\\
f^{\mu\nu}_{\:\:\:\:\:;\nu} &=& 0 \label{EM1N}\\
\ast f^{\mu\nu}_{\:\:\:\:\:;\nu} &=& 0 \label{EM2N}\\
f^{k\mu\nu}_{\:\:\:\:\:\:\:\:\mid \nu} &=& 0 \label{ymvfe1N}\\
\ast f^{k\mu\nu}_{\:\:\:\:\:\:\:\:\mid \nu} &=& 0 \ . \label{ymvfe2N}
\end{eqnarray}

The field equations (\ref{EM1N}-\ref{EM2N}) provide two electromagnetic potentials \cite{CF}, not independent from each other, but due to the symmetry of the equations, available for our construction. $A^{\mu}$ and $\ast A^{\mu}$ are the two electromagnetic potentials, see the comments made about the Abelian potentials and the star nomenclature $\ast A^{\mu}$ in section \ref{overview}. Similar for the two Non-Abelian  equations (\ref{ymvfe1N}-\ref{ymvfe2N}). The Non-Abelian potential $A^{k\mu}$ is available for our construction as well \cite{RGLG,RG,JS,NC,MC,YM,RU,KI}. With all these elements put together, we can proceed to define the auxiliary antisymmetric field,

\begin{eqnarray}
\omega_{\mu\nu} = Tr(\ast\zeta_{\:\sigma\tau}\: \zeta_{\mu\nu} + \zeta_{\:\sigma\tau}\: \ast\zeta_{\mu\nu})\:Tr(\zeta^{\sigma\rho}_{\:\:\:\:\:\:\mid\rho}\:\ast \zeta^{\tau\lambda}_{\:\:\:\:\:\:\mid\lambda}) \ .\label{anti1}
\end{eqnarray}

This particular antisymmetric auxiliary field in our construction could also be alternatively chosen to be,

\begin{eqnarray}
\omega_{\mu\nu} =  Tr(\zeta_{\:\sigma\tau}\: \zeta_{\mu\nu})\:Tr(\zeta^{\sigma\rho}_{\:\:\:\:\:\:\mid\rho}\:\ast \zeta^{\tau\lambda}_{\:\:\:\:\:\:\mid\lambda})\ .\label{anti2}
\end{eqnarray}

We can choose this antisymmetric auxiliary field $\omega_{\mu\nu}$ in many different ways, we just show two examples. It is clear that (\ref{anti1}) or (\ref{anti2}) are invariant under $SU(2)$ local gauge transformations. Expressions (\ref{anti1}) or (\ref{anti2}) are nothing but explicit examples among many, see for example reference \cite{A2}. Once our choice is made we perform a local duality rotation in order to obtain the new extremal field. We remind ourselves through the algorithm created in section \ref{overview} and reference \cite{A} that extremal fields are found through local duality rotations of second rank antisymmetric tensors like in equation (\ref{drefn}) because then we can use the equations analogous to (\ref{ic}-\ref{id}) to define an orthogonal tetrad. That is the core of this algorithm.

\begin{eqnarray}
\epsilon_{\mu\nu} = \cos\vartheta \: \omega_{\mu\nu} - \sin\vartheta \:\ast \omega_{\mu\nu}\ .\label{extremalRn}
\end{eqnarray}

As always we choose this complexion $\vartheta$ to be defined by the condition,

\begin{eqnarray}
\epsilon_{\mu\nu}\:\ast \epsilon^{\mu\nu} &=& 0\ .\label{rcn}
\end{eqnarray}

Thus we find the new local scalar complexion analogously to section \ref{overview} to be,

\begin{eqnarray}
\tan(2\vartheta) = - \omega_{\mu\nu}\:\ast \omega^{\mu\nu} / \omega_{\lambda\rho}\:\omega^{\lambda\rho}\ .\label{compr}
\end{eqnarray}

We used our new algorithm to find a new kind of local $SU(2)$ gauge invariant extremal tensor $\epsilon_{\mu\nu}$, that enables the construction of the new tetrad,

\begin{eqnarray}
S_{(1)}^{\mu} &=& \epsilon^{\mu\lambda}\:\epsilon_{\rho\lambda}\:X^{\rho}
\label{S1N}\\
S_{(2)}^{\mu} &=& \sqrt{-Q_{ym}/2} \: \epsilon^{\mu\lambda} \: X_{\lambda}
\label{S2N}\\
S_{(3)}^{\mu} &=& \sqrt{-Q_{ym}/2} \: \ast \epsilon^{\mu\lambda} \: Y_{\lambda}
\label{S3N}\\
S_{(4)}^{\mu} &=& \ast \epsilon^{\mu\lambda}\: \ast\epsilon_{\rho\lambda}
\:Y^{\rho}\ ,\label{S4N}
\end{eqnarray}

where $Q_{ym} = \epsilon_{\mu\nu}\:\epsilon^{\mu\nu}$ and we assume this local scalar not to be zero. With the help of identity (\ref{ib}), when applied to the case $A_{\mu\alpha} = \epsilon_{\mu\alpha}$ and $B^{\nu\alpha} = \ast \epsilon^{\nu\alpha}$ we obtain as in section \ref{overview} the equivalent condition to (\ref{rcn}),

\begin{eqnarray}
\epsilon_{\alpha\nu}\:\ast \epsilon^{\mu\nu} &=& 0\ ,\label{isu2n}
\end{eqnarray}

It is a simple excercise using (\ref{ib}) for $A_{\mu\alpha} = \epsilon_{\mu\alpha}$ and $B^{\nu\alpha} = \epsilon^{\nu\alpha}$, and (\ref{isu2n}), to prove that the vectors (\ref{S1N}-\ref{S4N}) are orthogonal. As we did before in section \ref{overview} we will call for future reference $\epsilon^{\mu\lambda}\:\epsilon_{\rho\lambda}$ the skeleton of the tetrad vector $S_{(1)}^{\mu}$, and $X^{\rho}$ the gauge vector. In the case of $S_{(3)}^{\mu}$, the skeleton is $\ast \epsilon^{\mu\lambda}$, and $Y_{\lambda}$ the gauge vector. It is clear now that skeletons are gauge invariant under $SU(2) \times U(1)$ as we announced at the start of this section. This property guarantees that the vectors under local $U(1)$ or $SU(2)$ gauge transformations will not leave their original planes or blades, keeping therefore the metric tensor explicitly invariant. Our final task in this construction will be to define the gauge vectors $X^{\sigma}$ and $Y^{\sigma}$ for the tetrad (\ref{S1N}-\ref{S4N}). A non-trivial although useful choice that we can make is $X^{\sigma} = Y^{\sigma} = Tr[\Sigma^{\alpha\beta}\:E_{\alpha}^{\:\:\rho}\: E_{\beta}^{\:\:\lambda}\:\ast \xi_{\rho}^{\:\:\sigma}\:\ast \xi_{\lambda\tau}\:A^{\tau}]$. The nature of the object $\Sigma^{\alpha\beta}$ is explained in section VI, Appendix II in reference \cite{A2} and also section \ref{sec:appI}. The object $\Sigma^{\alpha\beta}$ is basically built with the Pauli matrices and the identity two by two. The tetrad vectors $E_{\alpha}^{\:\:\rho}$ inside the expression $Tr[\Sigma^{\alpha\beta}\:E_{\alpha}^{\:\:\rho}\: E_{\beta}^{\:\:\lambda}\:\ast \xi_{\rho}^{\:\:\sigma}\:\ast \xi_{\lambda\tau}\:A^{\tau}]$ can be chosen to be the tetrad vectors that we already know from manuscript \cite{A} and section \ref{overview} for electromagnetic fields in curved space-times. Following the same notation as in \cite{A} and equations (\ref{UON}-\ref{WON}), we call $E_{(o)}^{\:\:\rho} = U^{\rho}$, $E_{(1)}^{\:\:\rho} = V^{\rho}$, $E_{(2)}^{\:\:\rho} = Z^{\rho}$, $E_{(3)}^{\:\:\rho} = W^{\rho}$. The electromagnetic extremal tensor $\xi_{\rho\sigma}$, and its dual $\ast \xi_{\rho\sigma}$ are also already known from reference \cite{A} and section \ref{overview}. We make use of the already defined tetrads built for Einstein-Maxwell spacetimes in order to enable the use of the object $\Sigma^{\alpha\beta}$ which is key in our construction. The key lies in the translating property of this object between $SU(2)$ local gauge transformations $S$ and local Lorentz transformations $\Lambda^{\alpha}_{\:\:\:\gamma}$, see reference \cite{A2} and notice from section \ref{sec:appI} that $S^{-1}\:\Sigma^{\alpha\beta}\:S = \Lambda^{\alpha}_{\:\:\:\gamma}\:\Lambda^{\beta}_{\:\:\:\delta}\:\Sigma^{\gamma\delta}$. We would like to study one more property of these chosen gauge vector fields $X^{\sigma} = Y^{\sigma} = Tr[\Sigma^{\alpha\beta}\:E_{\alpha}^{\:\:\rho}\: E_{\beta}^{\:\:\lambda}\:\ast \xi_{\rho}^{\:\:\sigma}\:\ast \xi_{\lambda\tau}\:A^{\tau}]$. The structure $E_{\alpha}^{\:\:[\rho}\:E_{\beta}^{\:\:\lambda]}\:\ast \xi_{\rho\sigma}\:\ast \xi_{\lambda\tau}$ is invariant under $U(1)$ local gauge transformations. The electromagnetic extremal field property \cite{A}$^{,}$\cite{MW}, $\xi_{\mu\sigma}\:\ast \xi^{\mu\tau} = 0$ is useful in the contractions $E_{\alpha}^{\:\:\rho}\: E_{\beta}^{\:\:\lambda}\:\ast \xi_{\rho}^{\:\:\sigma}\:\ast \xi_{\lambda\tau}$. Because it is leaving in the contraction of $E_{\alpha}^{\:\:\rho}\: E_{\beta}^{\:\:\lambda}$ with $\ast \xi_{\rho\sigma}\:\ast \xi_{\lambda\tau}$ only the antisymmetric object $E_{2}^{\:\:[\rho}\:E_{3}^{\:\:\lambda]}$, which is locally $U(1)$ gauge invariant. Precisely because of property (\ref{id}). Let us remember that the object $\Sigma^{\alpha\beta}$ is antisymmetric and contracted with the electromagnetic tetrads as $\Sigma^{\alpha\beta}\:E_{\alpha}^{\:\:\rho}\: E_{\beta}^{\:\:\lambda}$ inside the local gauge vector, see section \ref{nonabeltetrads}.

In the first paper \cite{A} we proved that the group $U(1)$ is isomorphic to the local group of boosts plus discrete transformations on blade one that we called LB1. The same group $U(1)$ is isomorphic to $SO(2)$, that we also called LB2 since it is related to local tetrad rotations on blade two. This is a fundamental result in group theory alone, let alone in physics. We proved in references \cite{A2,A3} that the local group of $SU(2)$ gauge transformations is isomorphic to the tensor product of three LB1 groups. Second, the local group of $SU(2)$ gauge transformations is isomorphic to the tensor product of three LB2 or $SO(2)$ groups. All the local gauge groups of the Standard Model have been proven to be isomorphic to local groups of tetrad transformations in four-dimensional Lorentzian curved or flat spacetimes. The no-go theorems of the sixties \cite{SWNG,LORNG,CMNG} have been proven to be incorrect. Not because of their internal logic but for the assumptions made at the outset of these theorems. We read in reference \cite{CMNG} ``S (the scattering matrix) is said to be Lorentz-invariant if it possesses a symmetry group locally isomorphic to the Poincar\`{e} group P.\ldots A symmetry transformation is said to be an internal symmetry transformation if it commutes with P. This implies that it acts only on particle-type indices, and has no matrix elements between particles of different four-momentum or different spin. A group composed of such transformations is called an internal symmetry group''. The local electromagnetic gauge group of transformations $U(1)$ has been proven to be isomorphic to local groups of tetrad transformations LB1 and LB2 on both the orthogonal planes one and two. These local groups of transformations LB1 and LB2$=SO(2)$ are composed of Lorentz transformations except in LB1 for an improper discrete reflection, see reference \cite{A}. Therefore the local Lorentz group of spacetime transformations cannot commute with LB1 or LB2 since Lorentz transformations on a local plane do not necessarily commute with Lorentz transformations on another local plane at the same point in spacetime. The local internal groups of transformations do not necessarily commute with the local Lorentz transformations, because they are isomorphic to local groups of tetrad transformations. Analogous results were proven for the non-Abelian cases $SU(2) \times U(1)$ and $SU(3) \times SU(2) \times U(1)$ Yang-Mills, see references \cite{A2,A3,ASU3}.

\section{Spacetime Feynman Calculus}
\label{spacetimefeynman}

It is of fundamental importance to understand the geometry of spacetime when particle interactions are taking place. Using the accumulated analysis for different kinds of gauge theories carried out in \cite{A,IWCP,A2,A3,ASU3,dsmg,gaugeinvmeth}, we will show explicitly how to assign to different Feynman diagrams in weakly interacting processes, different sets of tetrad vectors. The massive weak interactions boson mediators have an associated gravitational field as well as electrons, muons and neutrinos and even though these gravitational fields might be weak, they possess the necessary geometrical structure that enables the local symmetries of the standard model to be realized in an explicit fashion as it was analyzed in previous manuscripts \cite{A,IWCP,A2,A3,ASU3}. We judge relevant to understand that the transformations of colliding particles into the same or other emerging particles can occur through the local transformation properties of gravitational fields which will differ for different settings even though they exhibit analogous manifest local invariance under the internal symmetries of the standard model. We remind ourselves that in manuscripts \cite{A,IWCP,A2,A3,ASU3,dsmg,gaugeinvmeth} all the local internal gauge symmetries of the standard model have been proved isomorphic to local groups of tetrad transformations in four-dimensional curved Lorentz spacetimes. However, in order not to get foundational contradictions with quantum field theories at this stage of analysis we will assume that the kind of tetrads introduced in sections \ref{overview} and \ref{nonabeltetrads} are defined in Minkowski spacetime. This choice of spacetime involves no contradiction since the basic operation of local duality field transformation can be performed in flat Minkowski spacetimes as well. We can define the tetrads in Minkowski spacetime and prove that these tetrads define locally two orthogonal planes of stress-energy diagonalization. These geometrical structures exist not only in curved spacetimes but also in flat spacetimes. In essence local tetrad gauge states of spacetime would represent different microparticles in their spacetime manifestation, that can transform through local gauge transformations into other microparticles or other local tetrad gauge states of spacetime. The notation is a replica of the notation in \cite{DG}, so we refer the reader to this reference. We also refer the reader to \cite{DG,CL,GR,GMS,GTH} for abundant literature citation, specially in the field of particle physics.

\subsection{Weak interactions}

The existence of mediators as it was shown in \cite{A,IWCP,A2,A3} is irreplaceable as far as we are concerned with the construction of these kind of tetrads in weak interactions. In this case it is the existence of local $SU(2)$ ``extremal'' fields that allow us to build tetrads in weak processes. There are interactions involving the massive mediators where any virtual effect is negligible. For instance the $W^{-}$ as the mediator in inverse Muon decay. The $Z^{o}$ mediator in elastic Neutrino-Electron scattering. This is important because the existence of virtual processes would require a different approach. We will analyze these processes through the use of appropriately defined tetrads.

\subsubsection{Inverse Muon decay}
\label{invmuon}

Let us consider the process $e^{-}(1) + \nu_{\mu}(2) \rightarrow \nu_{e}(3) + \mu^{-}(4)$. There are two vertices. We invoke then the existence of the $SU(2)$ tetrads introduced in \cite{A2,A3}, specially the general tetrad structure presented in the section ``Gauge geometry'' and also section \ref{nonabeltetrads}. We called these general $SU(2)$ tetrad vectors $S_{(1)}^{\mu} \cdots S_{(4)}^{\mu}$ and the structure of these latter tetrads was introduced in equations (\ref{S1N}-\ref{S4N}) in the section \ref{nonabeltetrads} dedicated to the overview of these objects.

%\begin{eqnarray}
%S_{(1)}^{\mu} &=& \epsilon^{\mu\lambda}\:\epsilon_{\rho\lambda}\:X^{\rho}
%\label{S1IMD}\\
%S_{(2)}^{\mu} &=& \sqrt{-Q_{ym}/2} \: \epsilon^{\mu\lambda} \: X_{\lambda}
%\label{S2IMD}\\
%S_{(3)}^{\mu} &=& \sqrt{-Q_{ym}/2} \: \ast \epsilon^{\mu\lambda} \: Y_{\lambda}
%\label{S3IMD}\\
%S_{(4)}^{\mu} &=& \ast \epsilon^{\mu\lambda}\: \ast\epsilon_{\rho\lambda}
%\:Y^{\rho}\ ,\label{S4IMD}
%\end{eqnarray}

%where $\epsilon_{\mu\nu}$, is a local $SU(2)$ gauge invariant extremal tensor, and $Q_{ym} = \epsilon_{\mu\nu}\:\epsilon^{\mu\nu}$ is assumed not to be zero \cite{A2,A3}.
%We remind ourselves once again that in order not to enter at this stage in contradictions with the foundations of quantum field theory, the background spacetime is considered to be Minkowski.
%A treatment for a curved spacetime where a gravitational field is present would entail several new notions that we would like not to introduce at this stage of analysis.
There was a remaining freedom in the choice of two vector fields, $X^{\rho}$ and $Y^{\rho}$. It is exactly through an appropriate choice for these two vector fields that we can identify a tetrad set for each vertex at the same spacetime point. In addition to the previously introduced notation and structures, let us call the non-null electromagnetic tetrads, following again the notation in references \cite{A,IWCP,A2,A3} and sections \ref{overview}-\ref{nonabeltetrads}, $E_{\alpha}^{\:\:\rho}$. There are local non-null electromagnetic tetrads in both vertices at the same spacetime point since in one vertex we have an electron and in the other vertex a muon. The indices $\alpha$ and $\beta$ are reserved for locally inertial coordinate systems. Then, we can proceed to define for the first vertex the two gauge vector fields,

\begin{eqnarray}
X^{\rho} = Y^{\rho} = \overline{u}(3)\: {\bf \gamma^{\alpha}\: (1-\gamma^{5})}\:u(1)\:E_{\alpha}^{\rho} \ . \label{VERTEX1}
\end{eqnarray}

We are basically associating to the first vertex a current \cite{DG} $j^{\alpha}_{-}= \overline{u}(3)\: {\bf \gamma^{\alpha}\: (1-\gamma^{5})}\:u(1)$. This current describes the process $e^{-} \rightarrow \nu_{e} + W^{-}$. For the second vertex we can choose for instance,

\begin{eqnarray}
X^{\rho} = Y^{\rho} = \overline{u}(4)\: {\bf \gamma^{\alpha}\: (1-\gamma^{5})}\:u(2)\:E_{\alpha}^{\rho} \ . \label{VERTEX2}
\end{eqnarray}

Again, we are assigning to the second vertex a current \cite{DG} $j^{\alpha}_{-}= \overline{u}(4)\: {\bf \gamma^{\alpha}\: (1-\gamma^{5})}\:u(2)$ describing the process $\nu_{\mu} + W^{-} \rightarrow \mu^{-}$. It is evident from all the analysis in \cite{A2,A3} that the geometrical transition from vertex one to vertex two and vice-versa, is an $SU(2)$ generated local gauge transformation. That is only allowed through the existence of massive mediators. Following the ideas in \cite{A2} we can start by choosing for instance,

\begin{eqnarray}
X^{\rho} = Y^{\rho} &=& Tr[\Sigma^{\alpha\beta}\:E_{\alpha}^{\:\:\sigma}\: E_{\beta}^{\:\:\lambda}\:\ast \xi_{\sigma}^{\:\:\rho}\:\ast \xi_{\lambda\tau}\:A^{\tau}]  \label{startgtb1}
%Y_{\rho} &=& Tr[\Sigma^{\alpha\beta}\:E_{\alpha}^{\:\:\sigma}\: E_{\beta}^{\:\:\lambda}\:\ast \xi_{\sigma\rho}\:\ast \xi_{\lambda\tau}\:A^{\tau}]\ . \label{startgtb2}
\end{eqnarray}

The $\Sigma^{\alpha\beta}$ objects are analyzed in appendix II in reference \cite{A2} and sections \ref{nonabeltetrads}-\ref{sec:appI} in this present paper, $\xi_{\sigma\rho}$ are the electromagnetic ``extremal'' fields introduced in reference \cite{A} and section \ref{overview} in this present paper, etc. Through a local $SU(2)$ gauge transformation on blade one, we can ``rotate'' the normalized version of vectors (\ref{S1N}-\ref{S2N}) on blade one, until $X^{\rho}$ in (\ref{startgtb1}) becomes $X^{\rho}$ in (\ref{VERTEX1}). We can also ``rotate'' the normalized version of vectors (\ref{S3N}-\ref{S4N}) on blade two, until $Y^{\rho}$ in (\ref{startgtb1}) becomes $Y^{\rho}$ in (\ref{VERTEX1}). Let us remember that the tetrad skeletons are locally gauge invariant under $U(1) \times SU(2)$. This is just a sample of local gauge transformations of the normalized version of vectors (\ref{S1N}-\ref{S4N}). We proved in references \cite{A,A2,A3} that the maps that both in the local plane one and two send the tetrad vectors that generate these planes from an initial gauge vector into another gauge vector are injective and surjective maps. The map that assigns local groups of gauge transformations into local groups of tetrad transformations on either local orthogonal planes of stress-energy symmetry are isomorphisms. Again we can start with (\ref{startgtb1}) and appropriately ``rotate'' the tetrad vectors on blade one, until they become the ones corresponding to $X^{\rho}$ given in (\ref{VERTEX2}). Similar for $Y^{\rho}$ in this second case (\ref{VERTEX2}). It is evident then that (\ref{VERTEX1}) and (\ref{VERTEX2}) are connected through local $SU(2)$ gauge transformations on blades one and two, that in turn, leave invariant the metric tensor. That is, these local gauge transformations exist because of transitivity. The local groups of gauge transformations have been proven to be isomorphic to the local groups of tetrad transformations on the local orthogonal planes of symmetry. Given two sets of tetrads on the local plane one, then there is a local gauge transformation that sends one set into the other and vice-versa. Similar in the local othogonal plane two. These local orthogonal planes we remind ourselves are the local planes of covariant diagonalization of the stress-energy tensor, for the Abelian case and also for the non-Abelian case, see references \cite{A,IWCP,A2,A3,ASU3,gaugeinvmeth}.

We can also notice that the vector fields (\ref{VERTEX1}-\ref{VERTEX2}) are not strictly vectors but pseudovectors under local parity transformations, see reference \cite{DG}. But the metric tensor remains unaltered under these local parity transformations. It is as if the geometry associated to the $e^{-}(1)$ and $\nu_{e}(3)$ can be transformed through the existence of a massive mediator into the geometry associated to the $\nu_{\mu}(2)$ and $\mu^{-}(4)$ without altering the spacetime. The vertices are local tetrad gauge states of the same flat spacetime.

%\subsubsection{Decay of the Pion}

%Let us consider just one more example of a charged weak process, $\pi^{-}(1) %\rightarrow \overline{\nu}_{\ell}(2) + \ell^{-}(3)$, where $\ell$ is a muon or %an electron. We can choose then,

%\begin{eqnarray}
%X^{\rho} = Y^{\rho} = g_{w}\:\overline{u}(3)\: {\bf \gamma^{\alpha}\: (1-%\gamma^{5})}\:\upsilon(2)\:E_{\alpha}^{\rho} \ . \label{VERTEXPION}
%\end{eqnarray}

%Once more the vectors (\ref{VERTEXPION}) are pseudovectors under local parity %transformations, but again the metric tensor remains invariant under this kind %of transformation.

\subsubsection{Elastic Neutrino-Electron scattering}
\label{elasticneutrino}

Now, we are considering neutral currents. In particular the interaction process $\nu_{\mu}(1) + e^{-}(2) \rightarrow \nu_{\mu}(3) + e^{-}(4)$. As before we can assign to the first vertex the choice,

\begin{eqnarray}
X^{\rho} = Y^{\rho} = \overline{u}(3)\: {\bf \gamma^{\alpha}\: (1-\gamma^{5})}\:u(1)\:Z_{\alpha}^{\rho} \ . \label{VERTEX1NEL}
\end{eqnarray}

The current $j^{\alpha}_{-}= \overline{u}(3)\: {\bf \gamma^{\alpha}\: (1-\gamma^{5})}\:u(1)$, represents the process $\nu_{\mu}(1) \rightarrow \nu_{\mu}(3) + Z^{o}$. The tetrad $Z_{\alpha}^{\rho}$ is built as follows. Following again the notation in \cite{DG} we know we have available a local vector field $Z_{\mu}$ that results from the Weinberg rotation through the angle $\theta_{w}$, in addition to the standard electromagnetic local vector field $A_{\mu}$. The rotation can be written,

\begin{eqnarray}
A_{\mu} &=& B_{\mu}\:\cos\theta_{w} + W_{\mu}^{3}\:\sin\theta_{w}\\
Z_{\mu} &=& -B_{\mu}\:\sin\theta_{w} + W_{\mu}^{3}\:\cos\theta_{w}\ . \label{wr}
\end{eqnarray}

The local tetrad field $Z_{\alpha}^{\rho}$ is present in both vertices at the same spacetime point, since the massive neutral mediator is present in both local vertices and $Z_{\alpha}^{\rho}$ is a local tetrad associated to this flat spacetime. The electro-weak mixing involves a weak isotriplet of intermediate vector bosons $\bf{W}$ coupled to three weak isospin currents, and an isosinglet intermediate vector boson $B_{\mu}$ coupled to the weak hypercharge current. If we follow all the steps in reference \cite{A} and the method developed  in section \ref{overview} in the present paper, we can build out of the curl $Z_{\mu,\nu} - Z_{\nu,\mu}$ a new tetrad.  This auxiliary local tetrad $Z_{\alpha}^{\rho}$ present in the definition of gauge-vectors (\ref{VERTEX1NEL}) would once more in its own construction involve the choice of two gauge-vector fields, see reference \cite{A}. We can choose for instance $Z_{\mu}$ and $B_{\mu}$ as these two vector fields needed in turn in the definition and construction of the local auxiliary tetrad $Z_{\alpha}^{\rho}$. Then, the tetrad that couples to the neutrino current is associated to the massive $Z^{o}$.

The second vertex could be assigned a choice,

\begin{eqnarray}
X^{\rho} = Y^{\rho} = \overline{u}(4)\: {\bf \gamma^{\alpha}\: (c_{V}-c_{A}\:\gamma^{5})}\:u(2)\:E_{\alpha}^{\rho} \ , \label{VERTEX2NEL}
\end{eqnarray}

representing $e^{-}(2) + Z^{o} \rightarrow e^{-}(4)$. For this particular interaction $c_{V}=-{1 \over 2} + 2\:\sin\theta_{w}$, and $c_{A }=-{1 \over 2}$, where $\theta_{w}$ is again the Weinberg angle \cite{DG}. The massive mediator allows again for a $SU(2)$ local gauge transformation between the tetrad vectors chosen for vertex one and the ones chosen for vertex two at the same spacetime point. The neutral current works as a geometry mediator between the scattered particles keeping the spacetime invariant at the same spacetime point. The vertices function as local tetrad gauge states of the same spacetime.

%%%%%%%%%%%%%%%%%%%%%%%%%%%%%
%DISCUTIR s->c + W- | La matriz de Kobayashi-Maskawa.
%%%%%%%%%%%%%%%%%%%%%%%%%%%%%

\section{Conclusions}
\label{conclusions}

We have explored the possibility of assigning tetrads to Feynman diagrams. In interactions where we can assume the existence of particles with associated local fields like Abelian or non-Abelian gauge fields. At no step of analysis we have specified the tetrad themselves making all these geometrical properties outstanding since they can be put forward with all generality without the need to study case by case as long as the gauge fields are non-null for example. We can think the massive weak interactions boson mediators to have associated gravitational fields as well as electrons, muons and neutrinos do and even though these gravitational fields might be weak, they possess the necessary geometrical structure that enables the local symmetries of the standard model to be realized in an explicit fashion as it was studied thoroughly in previous manuscripts \cite{A,IWCP,A2,A3,ASU3}. However we decided to consider a background flat spacetime since gravitational fields would entail foundational contradictions with standard quantum field theories. New concepts would have to be introduced and we do not want to do this at this stage in analysis. We deem fundamental to understand that the transformations of colliding particles into the same or other emerging particles through elastic or inelastic processes can occur through the local transformation properties of spacetime which will differ for different settings through the new notion of tetrad gauge states of spacetime. Having done this explicitly, a number of questions naturally arise. We want these concluding remarks to be a summary of these open questions.

\begin{itemize}
\item The order of the formulations \cite{DG}$^{-}$\cite{RJ}. We have worked out the low-order diagrams. Then, what happens with higher order diagrams ?. The tetrads admit the choice of two gauge vector fields $X^{\rho}$ and $Y^{\rho}$, and the higher orders are additive exactly as in the quantum theories in these vector fields available as a choice. But there is more to understand. Do the higher order diagrams represent contributions coming from higher order perturbative theories of a full relativistic formulation of these interactions \cite{dsmg} involving perturbations of the electromagnetic field, etc, for instance, or just perturbative expansions in the gauge vector fields?. As an example of this kind of situation we might want to qualitatively consider the quark decays $b \rightarrow s\:\gamma$, see chapter XIV-7 in reference \cite{DGH} for instance. There are several possibilities but there are certainly higher order contributions to these kind of processes. Let us focus on the contribution that involves a $W^{-}$ boson mediator in rare decays. There is the $b \rightarrow W^{-} + c$ vertex and the subsequent $c + W^{-} \rightarrow s$ vertex for example. Each one of these has an associated current vector, let us call them for short $j^{\alpha}_{[bc]}$ and $j^{\alpha}_{[cs]}$. Both vertices have particles with electric charge so there is at each vertex an associated electromagnetic tetrad $E_{[bc]\alpha}^{\rho}$ and $E_{[cs]\alpha}^{\rho}$ respectively. Then we can associate to each vertex in this higher order diagram  gauge vectors $X_{[bc]}^{\rho} = Y_{[bc]}^{\rho} = j^{\alpha}_{[bc]}\:E_{[bc]\alpha}^{\rho}$ and $X_{[cs]}^{\rho} = Y_{[cs]}^{\rho} = j^{\alpha}_{[cs]}\:E_{[cs]\alpha}^{\rho}$. The $j^{\alpha}_{[bc]}\:E_{[bc]\alpha}^{\rho}$ contribution can then be added to the non-Abelian tetrad gauge vectors for vertex $[bc]$ and similar for $j^{\alpha}_{[cs]}\:E_{[cs]\alpha}^{\rho}$ at the vertex $[cs]$. In this contribution there is also a photon involved. This photon emission will change the stress-energy tensor and therefore will be associated to the perturbed extremal field which is in turn a local duality rotation of the perturbed electromagnetic field. These perturbation in the extremal field will in turn perturb the tetrad skeletons. Therefore, the vertices in the diagram will be associated to tetrad gauge states of the spacetime and the photon emission to tetrad skeleton perturbations. The $b \rightarrow s\:\gamma$ decays might involve contributions with top quarks and the analysis will be similar, the $b \rightarrow s\:\gamma$ decays could include a loop of the kind $c\overline{c}$ for example and we will proceed similarly as well. We will add to the gauge vectors associated to the corresponding vertex, higher and higher contributions with a corresponding expansion parameter. The currents at the corresponding vertex are the key objects necessary to produce gauge vectors associated to vertices. There could be that both, the perturbations in the skeletons and the perturbations in gauge vectors proceed simultaneously as in the $b \rightarrow s\:\gamma$ case. On one hand the local orthogonal planes of symmetry will tilt and on the other hand the tetrad vectors that span these local planes will rotate inside them.

\item A point outside the scope of this manuscript. The issue of ``gauge gravity''. Since in references \cite{A}$^{,}$\cite{A2} it was explicitly proved that the Abelian and non-Abelian gauge theories represent special symmetries of the gravitational field, we can ask about the meaning of ``gauge gravity''. The electromagnetic field is associated to the $LB1$ and $LB2$ symmetries of the gravitational field, see reference \cite{A}. The $SU(2)$ group of local gauge transformations is associated to the symmetries of the tensor product of three $LB1$ or three $LB2$ groups of transformations, see references \cite{A2,A3}. Analogous for $SU(3)$, see reference \cite{ASU3}. Then, it is not obvious to understand what is the meaning of a statement like, ``casting the theory of gravity into a Yang-Mills formulation''. We have reason to believe that we can truly cast the theory of gravity into a Yang-Mills formulation but as said, it is not obvious and requires a whole new work.

\item Another point outside the scope of this manuscript. The issue of quantum gravity. It has been proved explicitly that metric tensors can be associated with microparticle interactions. These constructions are possible by means of non-null Abelian tetrad fields, and by means of $SU(2)$ local non-Abelian tetrad fields. Perturbative formulations of these tetrad field structures as in reference \cite{dsmg} can take care of quantum fluctuations as well. The quantum is connected through the existence of these tetrad fields to gravity. A treatment for a curved spacetime where a gravitational field is present would entail several new notions that we would like not to introduce at this stage of analysis. Nonetheless we can advance that in curved spacetimes there will be local orthogonal planes one and two and the isomorphisms between local gauge groups Abelian and non-Abelian and local groups of tetrad transformations LB1 and LB2 would also apply in a similar fashion as to flat spacetimes \cite{A,IWCP,A2,A3,ASU3}. The quantization will be reflected through interactions that alter the local plane-symmetry structure by tilting these local planes of symmetry. Continuously for continuous perturbations and discretely in quantum settings situations. The main idea behind these quantum formulations in curved spacetimes is that the local planes of stress-energy diagonalization are always and during quantum interactions local orthogonal planes of symmetry, because quantum problems are basically confronted through perturbations analysis and these perturbations lead to continuous or discrete evolution of the local planes of symmetry either in flat or curved spacetimes. Continuous or discrete tilt symmetry evolution, see reference \cite{dsmg}. We also established that during interactions of microparticles the tetrad vectors that span the local orthogonal planes one and two might rotate inside them. The tetrads of different nature that we were able to build in \cite{A,IWCP,A2,A3,ASU3} and the present work, establish a link between the standard locally inertial flat field environment of the traditional standard quantum theories in weak interactions on one hand, and the curved spacetime of gravity on the other hand. The point is the following, why are we using in quantum gravity similar conceptual foundations to theories that are not formulated in curved spacetimes but flat spacetimes ?.
\item Another point outside the limited scope of this paper. The issue of the Higgs mechanism. It is a device conceived in its relationship with the nature of mass, for instance of the mass mediators. In the present tetrad environment we can ask if it is necessary, or the mass comes into existence due to the presence of gravity ?. Is it possible that the local Higgs field and its quantum fluctuations are related to the perturbations of the local gravitational weak field scalar approximation on a flat Minkowski background associated to the asymptotically curved spacetimes that in turn we can associate to elementary microparticles ?.
\item The issue of symmetry-breaking. It was proved in the general manuscripts \cite{A,IWCP,A2,A3} that the gravitational field when built with tetrads along the lines of expressions (\ref{S1N}-\ref{S4N}) are manifestly invariant under local electromagnetic gauge transformations, and local $SU(2)$ gauge transformations as well. But when assigning a tetrad set to a vertex in a low-energy weak process diagram in Minkowski spacetime, we make a particular choice for the two gauge vectors $X^{\rho}$ and $Y^{\rho}$. For instance, through associated currents we choose a particular gauge, and a different one for each vertex, like in inverse Muon decay or elastic Neutrino-Electron scattering. Then, we wonder if this gauge fixing procedure could be the geometrical form of the standard symmetry-breaking process. Hereby, we can see that it is the tetrad fields that bridge the two gauges associated to the two vertices, through a local $SU(2)$ gauge transformation, that in turn, leaves invariant the metric tensor.

\end{itemize}

\section{Appendix I}
\label{sec:appI}

This appendix is introducing the object $\Sigma^{\alpha\beta}$ that according to the matrix definitions introduced in the references is Hermitic. The use of this object in the construction of tetrads in section \ref{nonabeltetrads} enables the local $SU(2)$ gauge transformations $S$, to get in turn transformed into purely geometrical transformations. That is, local rotations of the $U(1)$ electromagnetic tetrads. The object $\sigma^{\alpha\beta}$ is defined \cite{MK}$^{,}$\cite{GM} as $\sigma^{\alpha\beta} = \sigma_{+}^{\alpha}\:\sigma_{-}^{\beta}-\sigma_{+}^{\beta}\:\sigma_{-}^{\alpha}$. The object $\sigma_{\pm}^{\alpha}$ arises when building the Weyl representation for left handed and right handed spinors. According to reference \cite{GM}, it is defined as $\sigma_{\pm}^{\alpha} = (\bf{1},\pm\sigma^{i})$, where $\sigma^{i}$ are the Pauli matrices for $i = 1\cdots3$. Under the $(\frac{1}{2},0)$ and $(0,\frac{1}{2})$ spinor representations of the Lorentz group this object transforms as,

\begin{equation}
S_{(1/2)}^{-1}\:\sigma_{\pm}^{\alpha}\:S_{(1/2)} = \Lambda^{\alpha}_{\:\:\:\gamma}\:\sigma_{\pm}^{\gamma}\ .\label{sigmatr}
\end{equation}

Equation (\ref{sigmatr}) means that under the spinor representation of the Lorentz group, $\sigma_{\pm}^{\alpha}$ transform as vectors. In (\ref{sigmatr}), the matrices $S_{(1/2)}$ are local objects, as well as \cite{GM} $\Lambda^{\alpha}_{\:\:\:\gamma}$. The $SU(2)$ elements can be considered to belong to the Weyl spinor representation of the Lorentz group. Since the group $SU(2)$ is homomorphic to $SO(3)$, they just represent local space rotations. It is also possible to define the object $\sigma^{\dagger\alpha\beta} = \sigma_{-}^{\alpha}\:\sigma_{+}^{\beta}-\sigma_{-}^{\beta}\:\sigma_{+}^{\alpha}$ in a similar fashion. Therefore, we can write,

\begin{center}
$\imath \: \left(\sigma^{\alpha\beta} + \sigma^{\dagger\alpha\beta}  \right)  = \left\{ \begin{array}{ll}
				0 \:\:\:\:\: \mbox{if $\alpha = 0$ and $\beta = i$}\\
				4\:\epsilon^{ijk}\:\sigma^{k} \:\:\:\:\: \mbox{if $\alpha = i$ and $\beta = j$ \ ,}
				    \end{array}
			    \right. $
\end{center}

\begin{center}
$ \sigma^{\alpha\beta} - \sigma^{\dagger\alpha\beta}  = \left\{ \begin{array}{ll}
				-4\:\sigma^{i} \:\:\:\:\: \mbox{if $\alpha = 0$ and $\beta = i$}\\
				0 \:\:\:\:\: \mbox{if $\alpha = i$ and $\beta = j$ \ .}
				    \end{array}
			    \right. $
\end{center}

We then may call $\Sigma_{ROT}^{\alpha\beta} = \imath \: \left(\sigma^{\alpha\beta} + \sigma^{\dagger\alpha\beta} \right)$, and $\Sigma_{BOOST}^{\alpha\beta} = \imath \: \left(\sigma^{\alpha\beta} - \sigma^{\dagger\alpha\beta} \right)$ and a possible choice for the object $\Sigma^{\alpha\beta}$ could be for instance $\Sigma^{\alpha\beta} = \Sigma_{ROT}^{\alpha\beta} + \Sigma_{BOOST}^{\alpha\beta}$. This is a good choice when we consider proper Lorentz transformations of the tetrad vectors nested within the structure of the gauge vectors $X^{\mu}$ and $Y^{\mu}$. For spatial rotations of the $U(1)$ electromagnetic tetrad vectors which in turn are nested within the structure of the two gauge vectors $X^{\mu}$ and $Y^{\mu}$, as is the case under study in section \ref{nonabeltetrads}, we can simply consider $\Sigma^{\alpha\beta} = \Sigma_{ROT}^{\alpha\beta}$. These possible choices also make sure the Hermiticity of gauge vectors.  Since when defining the gauge vectors $X^{\mu}$ and $Y^{\mu}$ we are taking the trace, then $X^{\mu}$ and $Y^{\mu}$ are real. 
%All the greek indices $\alpha$, $\beta$, $\delta$, $\epsilon$, $\gamma$, and $\kappa$, have been reserved in this manuscript for locally inertial coordinate systems. The importance of these objects $\Sigma^{\alpha\beta}$ is described in the section ``gauge geometry'' in paper \cite{A2} when we say:  ``We observe also the following. Since locally inertial transformations in general do not commute, then the locally $SU(2)$ generated transformations are non-Abelian. The non-Abelianity of $SU(2)$ is mirrored by the non-commutativity of these locally inertial transformations $\tilde{\Lambda}^{\alpha}_{\:\:\:\delta}$ of the electromagnetic tetrads. The key role in this non-commutativity is played by the object $\Sigma^{\alpha\beta}$, that translates local $SU(2)$ gauge transformations, into locally inertial Lorentz transformations''.

%\acknowledgements

%I am grateful to R. Gambini and J. Pullin for reading the first versions of this manuscript and for many fruitful discussions.

%\end{references}
\end{document}